\documentstyle[12pt,a4]{article}
\input psfig
\begin{document}
\normalbaselines
\addtolength{\topmargin}{-2cm}
\newcommand{\bm}{\bibitem}
\newcommand{\ud}{\bf}
\textheight=23cm
\thispagestyle{empty}
\renewcommand{\thefootnote}{\fnsymbol{footnote}}
\begin{center}
{\LARGE E2 component in subcoulomb breakup of $^{8}$B
\footnote{Work supported by EPSRC, UK, grant
no. GR/K33026.}}\\[1.0cm]
{\bf {R. Shyam$^{a}$\footnote{E-mail address: shyam@tnp.saha.ernet.in}
and I.J. Thompson$^{b}$}}\\ 
$^{a}${\it Saha Institute of Nuclear Physics, 
Calcutta - 700 064, India}\\[0.2 cm]
$^{b}${\it Department of Physics, University of Surrey, Guildford GU2 5XH,
U.K.}\\ 
\end{center}
\renewcommand{\thefootnote}{\arabic{footnote}}
\begin{abstract}
We calculate the angular distribution and total cross section of the
$^{7}$Be fragment emitted in the break up reaction of $^8$B on $^{58}$Ni
and $^{208}$Pb targets at the subCoulomb beam energy of 25.8 MeV, within
the non-relativistic theory of Coulomb excitation with proper three-body
kinematics. The relative contributions of the $E1$, $E2$ and $M1$ 
multipolarities to the cross sections are determined. 
The $E2$ component makes up about 65$\%$ and 40$\%$ of the $^7$Be total 
cross section for the $^{58}$Ni and $^{208}$Pb targets respectively. 
We find that the extraction of the astrophysical S-factor, $S_{17}(0)$,
for the $^{7}$Be(p,$\gamma$)$^8$B reaction at solar energies from the
measurements of the cross sections of the $^7$Be fragment in the Coulomb
dissociation of $^{8}$B at sub-Coulomb
energies is still not free from the uncertainties of the $E2$ component.

 
PACS NO. 25.70.De, 25.40.Lw, 96.60.Kx  
\end{abstract}
\newpage
The rate of the radiative capture reaction $^7$Be$(p,\gamma )^8$B 
at solar energies, is of considerable interest in the quest for
understanding the ``Solar neutrino puzzle".
The $^{37}$Cl  and Kamiokande detectors are particularly sensitive 
to the flux of the high energy neutrinos 
emitted in the subsequent $\beta$ decay of $^8$B [1]. Several attempts
have been made in the past to measure the rate of this reaction
at the lowest possible beam energies [2-7]. However, the measured cross
sections disagree in absolute magnitude, and the S-factors
($S_{17}(0)$) extracted
by extrapolating the data to solar energies ($\simeq$ 20 keV) differ
from each other by about 30-40 $\%$.

An alternate indirect way to determine the radiative capture
cross sections at low relative energies is provided by the Coulomb
dissociation method [8], which is based on the fact that the dissociation
of a projectile in the Coulomb field of a heavy target nucleus can be
considered as its photodisintegration. By using the principle of detailed
balance this cross section can be related to that of the radiative capture
process (we refer to Ref. [9] for a comprehensive review).

Motobayashi et al. have performed the first measurements of the
Coulomb dissociation of $^8$B into the $^7$Be$-p$ low
energy continuum in the field of $^{208}$Pb with a radioactive
$^8$B beam of 46.5 MeV/A energy [10]. We have presented earlier a detailed
analysis of this data [11], where the breakup cross sections of $^8$B
corresponding to $E1$, $E2$ and $M1$ transitions were calculated
using a theory of Coulomb excitation appropriate for intermediate beam
energies. Considering only the E1 component, the
measured breakup cross sections were
found to be consistent with a $S_{17}$(0) $\,=\,$(15.5 $\pm$ 2.80) eV barn,
which is smaller than even the lowest value reported by the direct
capture measurements. This is also appreciably smaller than the value
used in the standard solar model calculations [1]. However,
under the kinematical conditions of the experiment of Motobayashi et al.,
the $E2$ component of breakup may be disproportionately enhanced. In fact,
E2 corrections calculated from one of the models [12] of the structure of $^8$B
may lead to a further reduction of approximately 20 $\%$ in the value of
$S_{17}(0)$ [11,13]. Nevertheless, the contributions
of this component are strongly dependent on the model used to describe
the structure of $^8$B, and it is difficult to draw any definite
conclusion about the E2 contributions to this data [11,14,15]. This has led
to some of the authors of Ref. [10] to repeat this measurements with
angular distributions extended to larger scattering angles where the 
cross sections are expected to be more sensitive to the $E2$ component[16].

Recently von Schwarzenberg et al. [17] have measured
the breakup of $^8$B on the $^{58}$Ni target at the beam energy
of 25.8 MeV, well below the Coulomb barrier, where the $E2$ component of
the breakup is expected to be dominant. In contrast to the experiments
reported in Refs. [10,16]
where $^7$Be and $p$ were measured in coincidence, these
authors detect only the $^7$Be fragment. In their analysis of the data,
they have used the non-relativistic theory of Coulomb excitation [18]
and the radiative capture cross sections of Kim, Park and Kim (KPK) [12]
to estimate the $E1$, $E2$ and $M1$ component of the breakup cross
sections. However, the final state has been approximated as a
two body system by these authors. This implies that the measured angles of
$^7$Be were equated to those of the $^7$Be-$p$ center of mass (CM),
which may not be correct. Furthermore, the Coulomb excitation functions
needed in the calculations of the cross sections were obtained by
interpolating the values given in the tables of Ref. [18], which 
may lead to inaccuracies. 

In this letter, we present the results of an improved analysis of the data
of Ref. [16] by using a proper three body kinematics (TBK),
which avoids, automatically, equating the measured
$^7$Be angles with those of the CM of $^7$Be-$p$ system. Due to the 
difference in the masses of the two fragments, these two angles are
expected to be different. Furthermore, we use a proper three body
phase-space factor in the calculations of the cross sections.

In the first part of our presentation, we relate
the triple differential cross section for the Coulomb breakup of a
projectile (a) into its fragments (b and x) on a target A,
($A + a \rightarrow b + x + A$), to the cross section for the
Coulomb excitation (to the continuum) of the projectile a,
$A + a \rightarrow a^{*} + A$, which is calculated within the
Alder-Winther theory [18]. Using TBK
(see e.g. Ref. [19]), the momenta $\bf{p}_{bx}$ and $\bf{p}_{a^{*}}$
describing the relative motion of the fragments $b$ and $x$ and the
motion of their CM with respect to the target nucleus respectively,
can be related to their individual momenta $\bf{p}_b$ and $\bf{p}_x$
as following
\begin{eqnarray} 
{\mbox{\boldmath $ p $}_{bx}} & = & \mu_{bx}(\frac{{\mbox{\boldmath $ p $}}_b}
                                   {m_b} - \frac{{\mbox{\boldmath $ p $}}_x}
                                   {m_x} ) \\
{\mbox{\boldmath $ p $}_{a^{*}}} & = & {\mbox{\boldmath $ p $}}_{b}
                                    + {\mbox{\boldmath $ p $}}_{x}
                                    - \frac{m_{b}+m_{x}}{m_{a} + m_{A}}
                                      {\mbox{\boldmath $ P $}},
\end{eqnarray}
where
$m_{i}$ is the mass of the fragment $i$ and $\bf{P}$ is the
total momentum which is fixed by the conditions in the entrance channel.
$\mu_{bx}$ is the reduced mass of the $b-x$ system. Now let the
solid angles associated with the momenta $\bf{p}_b$, $\bf{p}_x$, $\bf{p}_{bx}$ and
$\bf{p}_{a^*}$ be $\Omega_b$, $\Omega_x$, $\Omega_{bx}$ and $\Omega_{a^*}$
respectively, then we can write
\begin{eqnarray}
\frac{d\sigma}{dE_{b}d\Omega_{b}d\Omega_{x}} & = & \frac{J}{4\pi}
                       \frac{d\sigma}{dE_{bx}d\Omega_{a^*}}
                         \frac{\partial E_x}{\partial E_{tot}},
\end{eqnarray}
where the total kinetic energy $E_{tot}$ is
\begin{eqnarray}
E_{tot} & = & E_b + E_x + E_A \\
        & = & E_{bx} + E_{a^{*}} + \frac{P^2}{2(m_a + m_A)},
\end{eqnarray}
is related to the projectile energy ($E_p$) and the reaction Q-value ($Q$) by
$E_{tot} \, = \, E_{p} + Q$. In Eq. (4) $E_b$, $E_x$, and $E_A$ are the
kinetic energies of the fragments $b$, $x$ and recoiling target nucleus
respectively, while $E_{bx}$ and $E_{a^{*}}$ are the kinetic energies of the
relative motion of the fragments and that of their CM with respect to the
target nucleus respectively. In Eq. (3), we have assumed
that the angular distribution of fragments is isotropic in the projectile
rest frame; the expressions without making this assumption are given in
Ref. [20]. The last factor in Eq. (3) is given by
\begin{eqnarray}
\frac{\partial E_x}{\partial E_{tot}} & = & m_A [m_x + m_A - m_x
                     \frac{{\mbox{\boldmath $ p $}}_{x} \cdot
                     ({\mbox{\boldmath $ P $}} - {\mbox{\boldmath $ p $}}_{b})}
                     {p_{x}^{2}}]^{-1},
\end{eqnarray}
and the Jacobian $J$ is defined as
\begin{eqnarray}
J & = & \frac{m_b p_b m_x p_x}{\mu_{bx} p_{bx} \mu_{aA} p_{a^{*}}} \ .
\end{eqnarray}
In Eq. (7) $\mu_{aA}$ is the reduced mass of the $a-A$ system.
The cross section $d\sigma/dE_{bx}d\Omega_{a^*}$ is related to the
photo-dissociation cross section as,
\begin{eqnarray}
\frac{d\sigma}{dE_{bx}d\Omega_{a^{*}}} & = & \frac{1}{E_{bx}}
                           \frac{dn_{\lambda}}{d\Omega_{a^{*}}}
                           \sigma(\gamma + a \rightarrow b + x),
\end{eqnarray}
where $dn_{\lambda}/d\Omega_{a^{*}}$ is the virtual photon number per unit
solid angle $\Omega_{a^{*}}$ for the relevant multipolarity ($\lambda$) in the
breakup process, and this can be calculated within the Alder-Winther theory.
The photodissociation cross section $\sigma(\gamma + a \rightarrow b + x)$ 
is related to the radiative capture cross section
$\sigma( b + x \rightarrow a + \gamma)$
by means of the detailed balance theorem. 

In Ref. [17], Eq. (8) has been used to get the total
cross section of $^7$Be by integrating this equation over the
relative energies $E_{bx}$ and the angular
aperture ($\pm$ 6$^\circ$) of the detectors (measuring $^{7}$Be)
placed at $45^\circ$ with respect to the beam direction. This procedure
necessarily assumes that the angles of $^7$Be are the same as those
$^8$B$^*$ ($^7$Be - $p$ CM). Such an assumption is avoided if this 
cross section is obtained by integrating Eq. (3) over the energy of 
$^7$Be and the solid angles ($\theta$, $\phi$) of the (unobserved) proton
and of $^7$Be. For given angles ($\theta_{{^7}Be}$, $\phi_{{^7}Be}$)
and ($\theta_p$, $\phi_p$), and energies $E_{^{7}Be}$ and $E_{^p}$, one can
use Eqs. (1) and (2) to determine the magnitude and directions of the
momenta $p_{^{7}Be-p}$ and $p_{^{8}B^{*}}$. In this way the cross sections
given by Eq. (3) can be determined from the Alder-Winther theory of
Coulomb excitation.

The angular distributions of $^8$B$^*$ are obtained by integrating
Eq. (8) over the relative energies $E_{bx}$. In our alternative approach, we
used the procedure outlined in the previous paragraph
to get the triple differential cross section
$d\sigma/d\Omega_{^{7}Be}dE_{^{7}Be}d\Omega_{p}$ and by integrating
them numerically over the solid angle $\Omega_{p}$ and
the energy $E_{{^7}Be}$, the angular distributions of the $^7$Be fragment
have been obtained. In Figs 1a and 1b, we show the angular distributions
(obtained by using the capture cross sections given by KPK model[12])
of $^8$B$^*$ and $^7$Be for the $^{58}$Ni target respectively. We note that 
the cross sections corresponding to the $E2$ component are larger than 
those of the other multipolarities in both cases (we have not shown here
the $M1$ components as they are very small). In Fig 1a, the $^8$B$^*$
cross sections for the $E1$ multipolarity are smaller than those of the
$E2$ at all the angles, which is in contrast to the results reported
in Ref. (17). Moreover, the magnitudes of both $E1$ and $E2$ components
are always larger than those given in this reference. This underlines
the inadequacy of the interpolation method used in Ref. [17] to obtain
the cross sections.
 
Although the angular distributions of the $^7$Be fragment look 
similar to those of $^8$B$^*$, the magnitude of the 
$E2$ component in the former case is about 10$\%$ larger
as compared to that in the latter case. The $E1$ component
is 15$\%$ smaller.

The ratio of the experimental total breakup cross section of $^7$Be
(obtained by integrating the breakup yields in the angular range,
($45 \pm 6$)$^\circ$, of the experimental setup) 
to Rutherford elastic scattering of $^7$Be is reported to be
$(8.1 \pm 0.8 \pm_{0.5}^{2.0}) \times 10^{-3}$, which is the only
quantity measured in Ref. [17]. It is not possible to get the total
breakup cross section of $^7$Be by directly 
integrating the angular distributions of various multipolarities
shown in Fig. 1a, as the corresponding angles belong to
$^8$B$^*$ and not to $^7$Be. Nevertheless, an approximate estimate
of this cross section can be made (from Eq. 8) by noticing that
the angles of $^7$Be can be related to those of
$^8$B$^*$ for given values of proton angles and magnitudes of the
$^7$Be and proton momenta (see Eqs. (1) and (2)). We find that
for the $^7$Be angles in the range of 39$^\circ$ - 51$^\circ$,
the $^8$B$^*$ angles vary between 41$^\circ$ to 65$^\circ$. We,
therefore, determine the the total cross section for the $^7$Be
fragment from Fig. 1a by summing the contributions of $E1$, $E2$
and $M1$ components which are obtained by integrating the corresponding
angular distribution over this angular range. The ratio of this cross
section to Rutherford elastic scattering of $^8$B is found to be
$ 6.8 \times 10^{-2}$. This is about an order of magnitude larger
than the experimental value. However, in the three-body case, where
the total cross section of the fragment $^7$Be   
can be obtained in a straight-forward way by integrating the 
distributions shown in Fig. 1b over the angular range of the experimental
setup, the value of this ratio is only $4.1 \times 10^{-2}$. 
It is interesting to note that the capture cross sections calculated
by some other authors [21,22,23] could lead to somewhat smaller values for
this ratio. For example, the model of Typel and Baur [21]
gives a value of $2.6 \times 10^{-2}$. Therefore, the values of this ratio 
predicted by various theoretical models of the capture reaction  
are larger than the upper limit of its reported experimental value by 
factors of 3 to 4. Thus the uncertainty about the magnitude of the
$E2$ cross section calculated in various models is not eliminated
by the measurement of Ref. [17], as this result is not reproduced by any
existing model of $^8$B. 

It may be remarked here that data for the angular distributions 
at larger angles may provide a better
regime for determining the $E2$ component in such an experiment. It can be
seen from Fig. 1b that beyond 50$^\circ$ the cross sections for this
component are about 3 to 5 times larger than the corresponding $E1$
cross sections. 

It is suggested in Ref. [17], that the method employed in their experiment
can be used to determine a precise value of $S_{17}$ if a heavy target is
used in the experiment and the data is taken at the same incident energy
and the angular range. This suggestion is, of course, based on the
assumption that for heavy targets the $E1$ component of breakup
would be predominant under the similar kinematical conditions.
We have examined the validity of
this assumption in Fig. 2, where we show the results of calculations
(performed using TBK) for $E1$ and $E2$ components of the reaction in
which  $^7$Be fragment is observed in the breakup of $^8$B on $^{208}$Pb
target at the same beam energy. Indeed, the $E1$ component is larger
than the $E2$ for certain angles. However, nowhere is the latter component
negligible; it even takes over the $E1$ component beyond $50^\circ$.
In the angular region of $30^\circ$ - $40^\circ$, where $E1$
component is large, $E2$ cross sections still contribute up to 40$\%$.
Therefore, no regime of the $^7$Be angular distribution is completely
free from the $E2$ component of the breakup, thus a clean determination
of the $S_{17}$ by performing
an experiment similar to that done in Ref. [17] on a heavy target appears
to be unlikely. In this regard, the experiments being carried out at
GSI at beam energies of 200 MeV/A are more promising as has already been
discussed in Ref. [11].

With the E2 contributions calculated with the TBK the dependence
of the fraction $f$ (($\sigma_{total}^{exp}$-$\sigma_{E1}^{calculated}$)/
$\sigma_{E2}^{calculated}$) on $S_{17}$ is different from that given in 
Ref [17]. The $f$ vs. $S_{17}$ curve looks more like that obtained
in the similar analysis of the data of Ref. [10] (the dashed-dotted curve 
in Fig. 5 of Ref. [17]) by Shyam et al. [11].

In summary, we have analysed the recently measured data on the breakup of
$^8$B on $^{58}$Ni target at the sub-Coulomb beam energy of 25.8 MeV.
In this experiment only $^7$Be fragment has been detected. We found that
with the proper three-body kinematics and phase-space factors
used in the calculations, the theoretical total cross sections are 
still larger than the upper limit of the experimental
data. Therefore, the present measurements do not completely eliminate the
uncertainty in the $E2$ predictions of various models proposed to
calculate the capture cross sections. Furthermore, the prospect of
determining a precise value for the astrophysical S-factor $S_{17}$
by performing the similar experiment with a heavy target does not seem
to be very encouraging, as in no angular regime is the $E2$ component of the
breakup negligible. A possible way to make the $E2$ component more
definite would be to measure the angular distributions of $^7$Be in
such an experiment, as those corresponding to $E1$ and $E2$ components are
quite different in shape as well as in magnitude, and can then be
easily separated from each other.

We would like to thank Jim Kolata for several useful discussions.

\newpage
\begin{center}{\bf Figure Captions} \end{center}
\begin{itemize}
\centerline{\psfig{figure=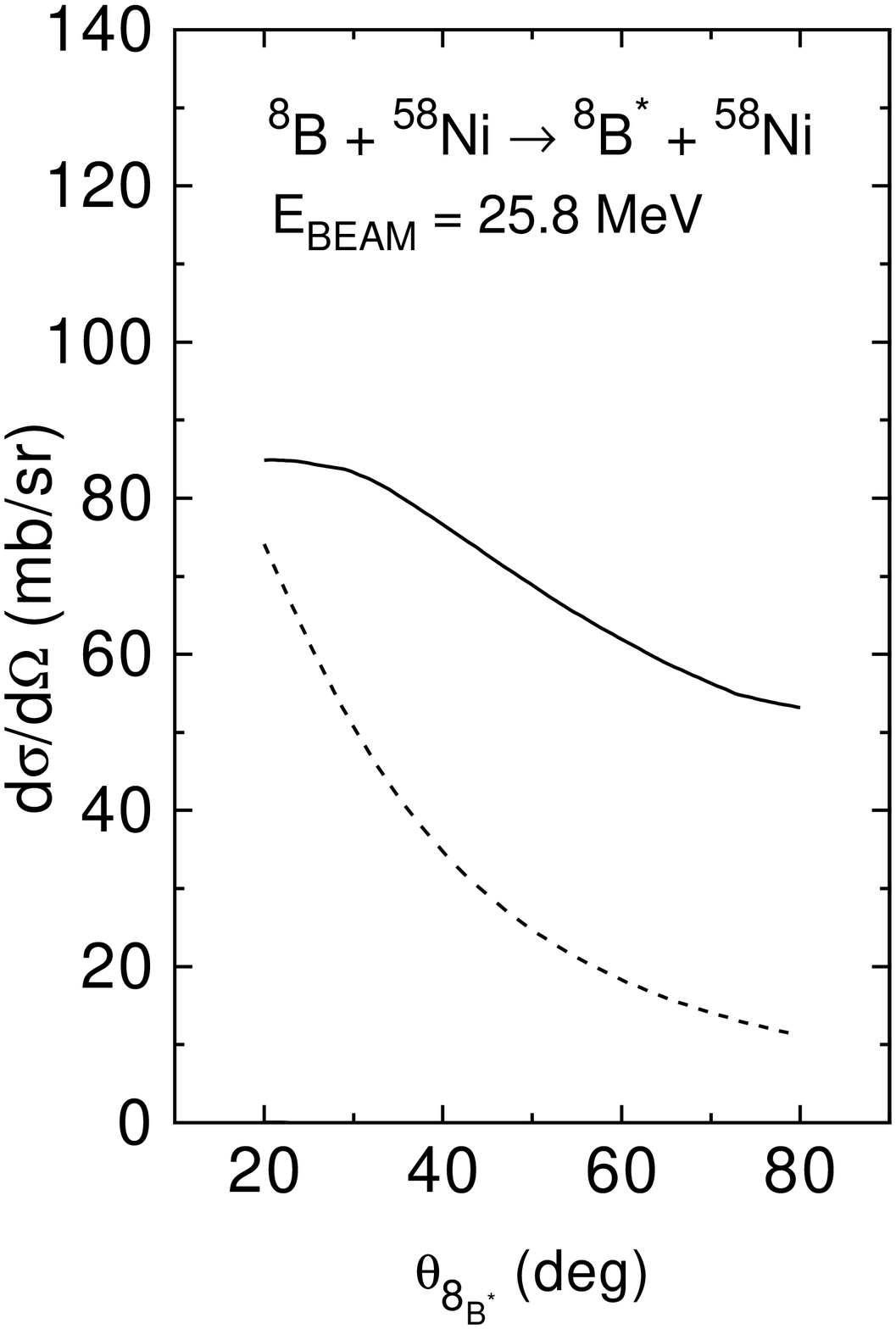,width=.9\textwidth}}
\item[Fig. 1a.] Angular distribution for $^8$B$^*$ in the Coulomb excitation
of $^8$B on  $^{58}$Ni target at the beam energy of 25.8 MeV.
The dashed and solid lines show the $E1$ and $E2$ cross sections respectively
which are obtained by using Eq (8).

\centerline{\psfig{figure=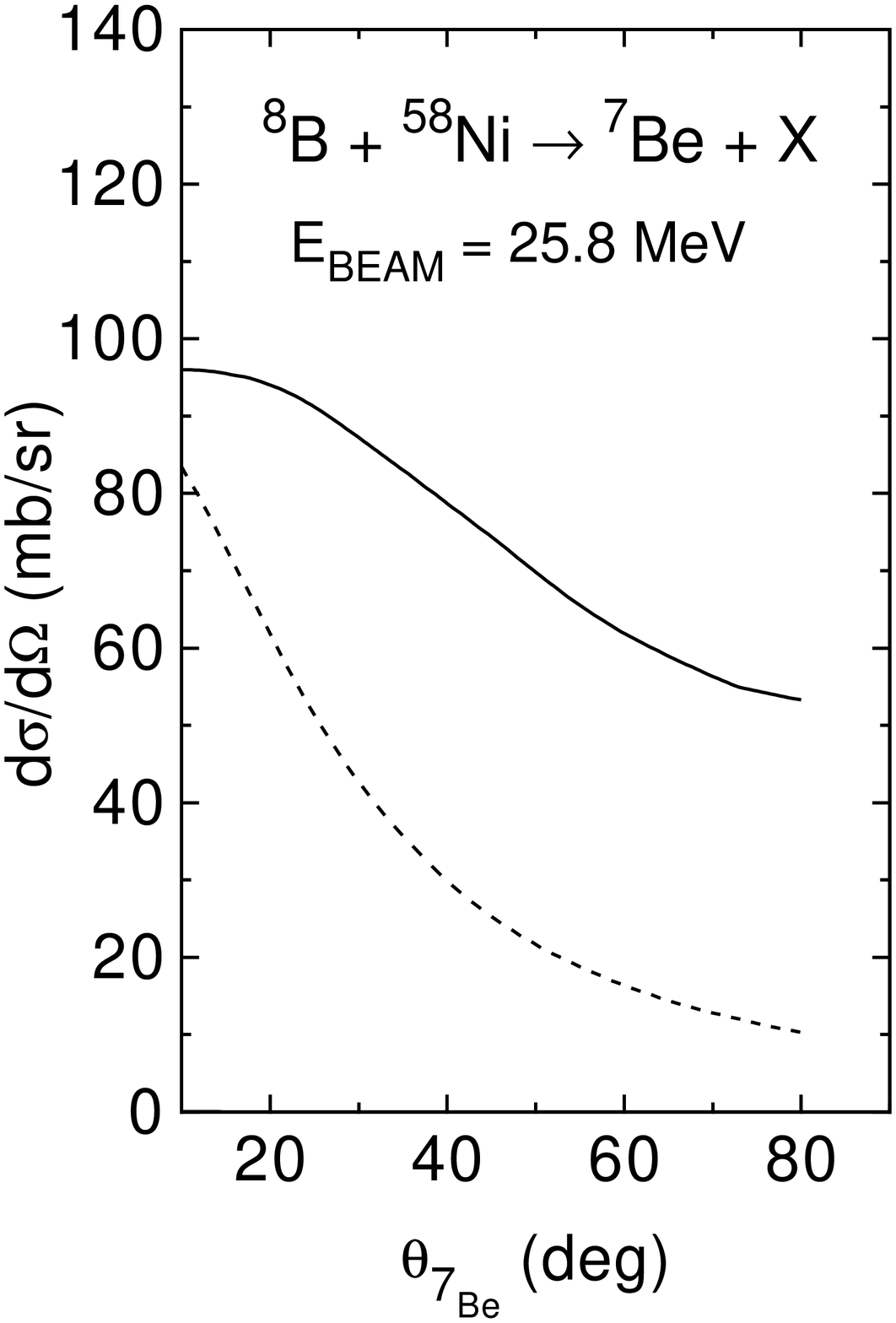,width=.9\textwidth}}
\item[Fig. 1b.] Angular distribution of the $^7$Be fragment emitted in the
breakup reaction of $^8$B on $^{58}$Ni target at the beam energy of 25.8 MeV.
The dashed and solid lines show the $E1$ and $E2$ cross sections respectively
which are calculated by using Eq. (3) with the proper three body kinematics. 

\centerline{\psfig{figure=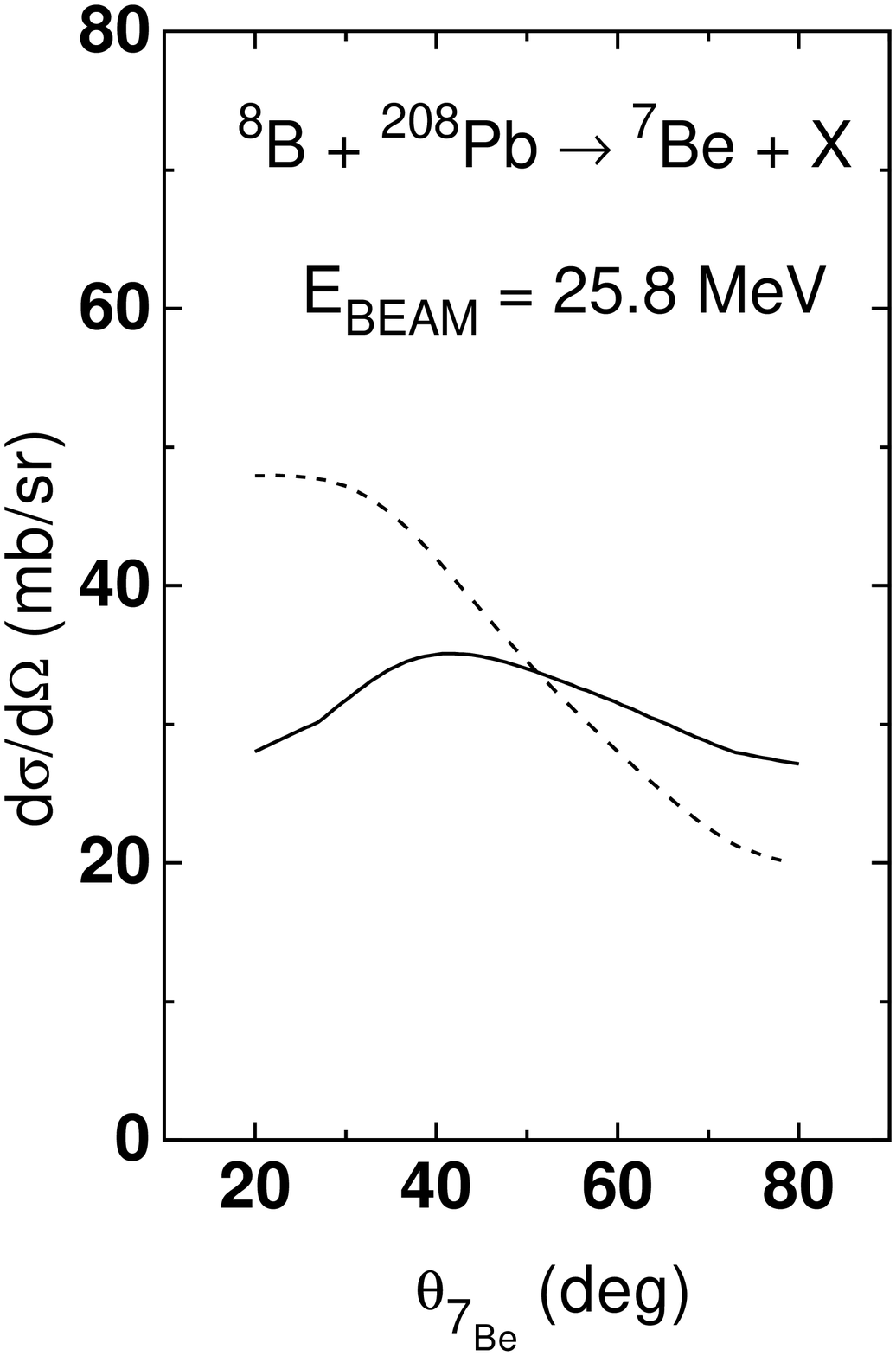,width=.9\textwidth}}
\item[Fig. 2.] Angular distributions of the $^7$Be fragment emitted in the
breakup of $^8$B on $^{208}$Pb target at the beam energy of 25.8 MeV. The
solid and dashed lines have the same meaning as in Fig. 1b.

\end{itemize} 
\end{document}